# Validation of a CT-brain analysis tool for measuring global cortical atrophy in older patient cohorts

Sukhdeep Bal,[1,4] Emma Colbourne,[1] Jasmine Gan,[1] Ludovica Griffanti,[2,3] Taylor Hanayik[2]

Nele Demeyere,[4] Jim Davies,[5] Sarah T Pendlebury,[1,6,7]* Mark Jenkinson[1,2] *

*Joint senior authors

[1]Wolfson Centre for Prevention of Stroke and Dementia, Wolfson Building, Nuffield Department of Clinical Neurosciences, University of Oxford, John Radcliffe Hospital, Headley Way, Oxford, OX3 9DU, UK.

[2]Centre for Integrative Neuroimaging, Nuffield Department of Clinical Neurosciences, University of Oxford, FMRIB Building, John Radcliffe Hospital, Headley Way, Headington, Oxford,
OX3 9DU, UK.

[3]Department of Psychiatry, University of Oxford, Warneford Hospital Oxford, Warneford Lane, Oxford, OX3 7JX, UK.

[4]Nuffield Department of Clinical Neurosciences, University of Oxford, Level 6, West Wing, John Radcliffe Hospital, Headley Way, Oxford, OX3 9DU, UK.

[5]Department of Computer Science, University of Oxford, Wolfson Building, Parks Road, Oxford, OX1 3QG, UK.

[6]Departments of Acute General (Internal) Medicine and Geratology, Oxford University Hospitals NHS Foundation Trust, John Radcliffe Hospital, Headley Way, Oxford, OX3 9DU, UK.

[7]NIHR Biomedical Research Centre, Oxford University Hospitals NHS Foundation Trust, John Radcliffe Hospital, Headley Way, Oxford, Oxfordshire OX3 9DU, UK.

Address correspondence to: Professor Sarah Pendlebury, Wolfson Centre for Prevention of Stroke and Dementia, Wolfson Building, John Radcliffe Hospital, Headley Way, Oxford OX3 9DU, UK. Email: sarah.pendlebury@ndcn.ox.ac.uk, Telephone: +44 1865 231603

**Abstract**

Background: Quantification of brain atrophy currently requires visual rating scales which are time consuming and automated brain image analysis is warranted.

Purpose: We validated our automated deep learning (DL) tool measuring the Global Cerebral Atrophy (GCA) score against trained human raters, and associations with age and cognitive impairment, in representative older ($\geq$65 years) patients.

Methods: CT-brain scans were obtained from patients in acute medicine (ORCHARD-EPR), acute stroke (OCS studies) and a legacy sample. Scans were divided in a 60/20/20 ratio for training, optimisation and testing. CT-images were assessed by two trained raters (rater-1=864 scans, rater-2=20 scans). Agreement between DL tool-predicted GCA scores (range 0-39) and the visual ratings was evaluated using mean absolute error (MAE) and Cohen's weighted kappa.

Results: Among 864 scans (ORCHARD-EPR=578, OCS=200, legacy scans=86), MAE between the DL tool and rater-1 GCA scores was 3.2 overall, 3.1 for ORCHARD-EPR, 3.3 for OCS and 2.6 for the legacy scans and half had DL-predicted GCA error between -2 and 2. Inter-rater agreement was *Kappa*=0.45 between the DL-tool and rater-1, and 0.41 between the tool and rater- 2 whereas it was lower at 0.28 for rater-1 and rater-2. There was no difference in GCA scores from the DL-tool and the two raters (one-way ANOVA, p=0.35) or in mean GCA scores between the DL-tool and rater-1 (paired t-test, t=-0.43, p=0.66), the tool and rater-2 (t=1.35, p=0.18) or between rater-1 and rater-2 (t=0.99, p=0.32). DL-tool GCA scores correlated with age and cognitive scores (both p<0.001).

Conclusion: Our DL CT-brain analysis tool measured GCA score accurately and without user input in real-world scans acquired from older patients. Our tool will enable extraction of standardised quantitative measures of atrophy at scale for use in health data research and will act as proof-of-concept towards a point-of-care clinically approved tool.

**Introduction**

Computed Tomography (CT) is the most widely used brain imaging modality globally owing to its low cost, tolerability and excellent sensitivity to acute haemorrhage (1). Around 60% of older acute hospital in-patients have CT-brain imaging either during admission or available from previous encounters with 80 million CT scans performed per year in the United States alone. Although MRI scanning provides better grey/white matter tissue contrast than CT, resource limitations and practical considerations substantially limit its use in older patients.

Routinely acquired CT-brain scans contain a wealth of information on brain ageing in the form of atrophy and white matter disease which predict risk of cognitive and functional decline (2-4). However, there are barriers in extracting standardised, quantitative information. Reporting of CT-brain images does not usually include detailed quantitative atrophy evaluation. Visual rating scales such as the Global Cortical Atrophy (GCA) scale or the medial temporal atrophy (MTA) scale are used infrequently and application is likely to vary between radiologists (1, 5, 6). Where the clinical indication is to exclude brain injury (e.g. after a fall) reporting may be brief and focussed on the immediate indication (7). Further, since radiology reports are in the form of free text, information is not easily extractable for populating electronic health records (EHRs) or automated algorithms (7). Standardised quantitative measurements of atrophy are therefore not routinely available as part of standard care.

Reliable automated brain image analysis of brain atrophy would therefore address a clinical and research need. Such a tool would need to be fully automated, requiring minimal or no user input, generalisable across different patient groups and datasets, and applicable to routinely acquired CT-brain scans for optimal clinical utility. However, despite the ubiquity of CT brain imaging, most atrophy measurement tools have been developed for MRI and the few existing CT brain analysis tools have limitations. Most are focussed on identifying acute pathology including stroke and those measuring atrophy provide volumetric outputs unfamiliar to clinicians.

We therefore developed an automated CT-brain analysis tool using novel AI Deep Learning (DL) methods for quantifying cerebral atrophy in the form of the GCA score in older patients (aged $\geq$65 years). In the current paper, we aimed to validate our CT-atrophy DL-tool by i) measuring differences in DL-predicted GCA score vs scores from trained human raters, ii) comparing DL-tool-to-human inter-rater variability with human-to-human inter-rater variability, iii) assessing DL-tool classification into one of three atrophy severity categories, iv) comparing DL-tool GCA scores with clinical measures including age and cognitive impairment.

**Methods**

*Patient Population*

We used routinely acquired CT-brain imaging from representative older patient cohorts from acute general (internal) medicine, acute stroke and a legacy set of older patient CT scans. The acute medicine cohort comprised consecutive patients admitted (2010-2018) at Oxford University Hospitals NHS Foundation Trust (OUHFT) incorporated into the Oxford Cognitive Comorbidity, Frailty and Ageing Research Database-Electronic Patient Records (ORCHARD-EPR, REC reference=18/SC/0184) (8, 9). The acute stroke scans were from the Oxford Cognitive Screen (OCS) stroke screening programme (10, 11) (REC references=14/LO/0648,18/SC/0550). Scans with no visible lesion were selected for the current study.

*Image Pre-processing*

Images were acquired on clinical CT scanners at the OUHFT as part of standard patient care. After image acquisition, CT brain images were de-identified and transferred from the Picture Archiving and Communication System (PACS) servers using an automated pipeline. CT brain images were skull stripped using the FSL-BET tool (12, 13) and linearly registered to a skull-stripped CT template (14).

*Application of the GCA scale to provide the reference standard*

The GCA scale (5) quantifies brain atrophy in 13 regions in total: 12 regions evaluated separately in each hemisphere (sulcal and ventricular dilatation) and one central region extending across both hemispheres (ventricular dilatation only). Sulcal dilation is assessed in the right and left hemispheres of the frontal, temporal and parieto-occipital lobes.

Ventricular dilation is assessed in the right and left hemispheres of the frontal, temporal and occipital horns, as well as the central third ventricle. Each of the 13 regions is assessed using a four-point scale (0=absent, 1=mild, 2=moderate and 3=severe atrophy) which is summed across all regions to give a total score out of 39 (5). For the current study, two raters operationalised the GCA to ensure consistent application across scans and regions (see Supplementary Methods). Rater-1 rated all study scans and Rater-2 rated a subset (n=20). Rater 1 recorded the time taken to apply the GCA scale in a consecutive set of 18 scans.

*Prediction of GCA scores using a DL model*

Pre-processed 3D CT brain images (n=864) were randomly separated (60/20/20 proportion) into training (n=518), optimisation (n=173) and testing (n=173) datasets. GCA scores from rater-1 were used as the ground truth during the DL-tool training. Training and optimisation were performed followed by tool testing using the testing set scored by rater-1. DL tool testing was also performed on the 20 scans rated by rater-2. Tool development was done using Python's pytorch library (15) on a workstation with hardware Intel I5/Ram:16 GB/GPU. DL-tool GCA score measurement took four seconds per scan.

*Statistical Analysis*

We first assessed DL tool performance in predicting the GCA scores in relation to the visually-rated GCA scores from rater-1. Analyses were undertaken first in the cohort as a whole (combining ORCHARD-EPR, OCS and legacy scans) and stratified by age $\geq$75 vs <75 years, and then for each of the three patient groups separately. Second, we assessed tool performance in predicting the GCA score in relation to scores from rater-2, to assess its generalisability across different raters. Third, we assessed the agreement of rater-1 vs rater-2 on the subset of 20 scans rated by both as a reference for human-to-human inter-rater variability. Fourth, we assessed the accuracy of classification into one of three severity groups as these are commonly used by clinicians according to the GCA scores (no/mild (GCA=0-11), moderate (GCA=12-21) and severe (GCA=22-39) atrophy) of the DL tool vs the GCA scores from rater-1. Finally, we correlated DL-tool GCA scores with age and cognitive scores (10-point Abbreviated Mental Test [AMT] <9 impaired and the Oxford Cognitive Screen) (10, 11).

Differences between the DL tool-predicted GCA and the visually rated scores were explored using scatter plots and Bland-Altman plots to give a measure of systematic bias. Mean absolute error (MAE) was used to assess the agreement between the DL tool-predicted GCA scores vs the visually rated GCA scores, separately for rater-1 and rater-2. We also assessed the differences between the GCA scores from the DL tool, rater-1 and rater-2 using a repeated-measures one-way ANOVA and also paired t-tests to check for differences in mean GCA scores between the DL-tool and rater-1, between the DL-tool and rater-2, and between rater-1 and rater-2.

We conducted an inter-rater assessment to examine the DL tool performance in comparison to the human raters and specifically whether the agreement between the DL tool and rater-1 was similar or better than between rater-1 and rater-2 since this would indicate that the DL-tool was able to obtain human level performance. Agreement between tool-predicted GCA vs the GCA from each of rater-1 and rater-2 was examined using Cohen's weighted Kappa with linear weighting and interpreted according to Landis and Koch (17). Confusion matrices were used to examine the accuracy of categorical severity classification (none/mild, moderate, severe) for the DL tool-predicted GCA vs the visually rated GCA scores from rater-1. Correlations with age and cognitive scores were performed using Spearman's Correlation. Differences in GCA scores between age groups (age : 65-75, age : 76-84, age > 84) were performed using Kruskal-Wallis test. Differences in GCA scores between age groups (age >75 and age <=75) and cognitively impaired and unimpaired groups were performed using an unpaired two-sample Wilcoxon test.

We used the Scikit-learn ([http://scikit-learn.org/stable/](http://scikit-learn.org/stable/)) ; Scipy ([https://www.scipy.org/](https://www.scipy.org/)) ; and NumPy ([http://www.numpy.org/](http://www.numpy.org/)) library to apply the statistics.

**Results**

Among 864 patients, 578 were from ORCHARD-EPR, 200 were OCS acute stroke patients and 86 were legacy patients. The mean age/SD for ORCHARD-EPR was 81/8 years, range=102-65 years (n=285 (49.3%) male) and for OCS cohorts was 78/8.4 years, range 60-95 years (n=88 (44%) male, Table 1). The distribution of age, and of GCA scores from rater-1, are shown in Figure 1. The average time to apply the GCA score, taken by rater-1 across the subset of 18 scans in which time taken was recorded, was 2 minutes and 56 seconds (Supplementary Table 1).

Scatter plots showed overall agreement of DL-predicted GCA scores with the visually-rated scores from rater-1 and this was the case for patients aged <75 and $\geq$75 years although the tool tended to underestimate high GCA scores (rater-1 GCA >28) and overestimate lower GCA scores (GCA <3, Figure 2A-D, Supplementary Figure 1A-B). The MAE for the difference between the DL-tool GCA scores and the GCA scores obtained by rater-1 was 3.2 for the dataset overall, 3.1 for ORCHARD-EPR, 3.3 for OCS, and 2.6 for the legacy scans (Figure 2A-D). For the subset of CT images (n=20) rated by both the raters, MAE was 3.5 between the DL-tool and rater-1, 4.7 between the DL- tool and rater-2 and 5.2 between rater-1 and rater-2 (Figure 2E,F). For 88 (50%) of the testing set CT scans, the GCA score error distribution for the DL-tool vs rater-1 was between -2 and 2 and for 132 (75%) of the testing scans, it was between -5 and +4 (Supplementary Figure 1C).

As visualised in the Bland–Altman plots (Figure 3), the mean of the signed differences in GCA scores estimated by the DL-tool vs those obtained by rater-1 for the combined cohort was 0.18 (limits of agreement, −7.8 to 8.2; Figure 3A). Across the subgroups, the mean of the differences was 1.44 (−6.6 to 9.5; Figure 3B) for ORCHARD-EPR, -0.88 (−8.3 to 6.6; Figure 3C) for OCS patients and 0.26 (−6.7 to 7.2; Figure 3D) for the legacy scans. No statistically significant difference was found in the GCA scores across the DL-tool and the two raters (F statistic=1.06, p=0.35, ANOVA) and the two-sample t-tests indicated no significant differences in mean GCA scores between the DL-tool and rater-1 (t=-0.43, p=0.66), between the DL-tool and rater-2 (t=1.35, p=0.18) or between rater-1 and rater-2 (t=0.99, p=0.32).

Inter-rater agreement was *Kappa*=0.45 for the DL-tool vs rater-1, 0.41 for the DL-tool vs rater-2, and 0.28 for rater-1 vs rater-2, indicating moderate agreement between the DL-tool-predicted GCA scores and either of the two raters, whereas agreement between the two raters was only fair. Regarding DL-tool classification into the three categories of none/mild, moderate and severe atrophy vs the rater-1 classifications, the accuracy of the DL-tool was 73% for mild atrophy and 70% for moderate and 70% for severe atrophy (Figure 4A). For DL-tool classification vs the rater-2 classifications, the accuracy of the DL-tool was 75% for mild atrophy, 69% for moderate and 33% severe atrophy (Figure 4B). For rater 1 vs the rater-2 classifications, the accuracy was 50% for mild atrophy, 77% for moderate and 33% severe atrophy (Figure 4C).

In ORCHARD-EPR, DL-tool GCA scores were correlated with both age ($\rho$=0.32, p=0.0003) and AMT scores ($\rho$ =-0.41, p=0.0001, Figure 5A,B). DL-tool mean/SD GCA was 14.4/3.9 for age <75 vs 17.2/4.2 for age $\geq$75 (Figure 5C) and increased across age categories: 13.9 (age 65-75), 16.9 (age 76-84), 17.6 (age > 84, p=0.003, Kruskal-Wallis). DL-GCA was higher in cognitively impaired (AMTS<9, mean/SD GCA=18.5/4, median=18) vs unimpaired (AMTS>=9, mean/SD GCA=15.4/4, median=15) patients (unpaired two-sample Wilcoxon test, p=0.003, Figure 5D). In the OCS cohort, the DL-GCA was also correlated with age ($\rho$=0.56, p<0.0001, Figure 6A). DL-tool mean/SD GCA was 9.5/2.9 vs 16.4/ 3.8 for age $\leq$75 vs >75 (p<0.0001, unpaired two-sample Wilcoxon test) (Figure 6C). However, the mean GCA scores were not significantly different between more cognitively impaired vs less cognitively impaired groups as defined by the median number of tasks impaired (11.9/4.6 vs 11.5/5.1, p=0.74, unpaired two-sample Wilcoxon test, Figure 6D).

**Discussion**

In this study, we validated our fully automated DL-based tool quantifying global brain atrophy in the form of a clinically meaningful GCA score on older patient CT-brain scans acquired in the course of standard clinical care. Our tool was reliable and fast, producing results in four seconds, and did not require any user input. The error in GCA scores obtained by the tool versus either of two trained human raters was similar or smaller than the difference between these same two human raters. The tool performed well across a wide range of GCA scores from minimal to severe atrophy, although it had a tendency to under-estimate high scores and over-estimate low scores.

Most available tools quantify brain atrophy through segmentation of brain structures providing outputs in the form of brain volumes (18-20). Widely used MRI segmentation methods have been adapted for CT (e.g. CTSeg) although tissues may be misclassified (18). In a recent study, U-Net models were used to segment brain volumes (ventricular CSF, grey matter, white matter) showing group level differences in patients with dementia vs cognitively healthy individuals (19). One previous tool measured GCA scores on CT scans (20). However, tool training was based on automated segmentation of paired MRI images, and all subjects had to have both CT and MRI done within six months. In addition, the GCA was quantified using a compressed scale of 0-3, most subjects had suspected cognitive decline and demographics were not described.

In developing our tool, we used an alternative approach: the DL-based tool was trained on large number of CT-brain images labelled with visually rated GCA scores without the need for tissue segmentation. Our tool quantified the GCA score which is familiar to clinicians and can be used to classify scans by the degree of atrophy. Tool performance was good relative to the visual ratings from rater-1 (MAE : 3 with no difference in scores by t-test, and kappa of 0.45) and was equivalent or better than between rater-1 and rater-2 (kappa of 0.28).

Similarly, the ability of the tool to classify the CT images as none/mild, moderate and severe atrophy, as defined by rater-1, was moderate and equivalent or better than that for rater-2 vs rater-1. Tool performance was similar in younger old patients (aged 65-75 years) and the older old (aged $\geq$75 years) and between different clinical groups (acute medicine and acute stroke patients).

Our DL-tool measures of atrophy correlated with age and also cognition although there was overlap in GCA scores between older and younger people and between cognitively impaired vs normal individuals as was also seen for segmentation volumes in previous studies (19). It should be noted that although associations between atrophy on brain imaging and cognition are well recognised, there is nevertheless considerable variation across individuals. In addition, the AMT is very brief and insensitive to milder cognitive impairments (21). Associations between tool measured GCA and cognition measured by the stroke-specific test in the OCS cohort were less strong than between the GCA and the AMT in the ORCHARD-EPR cohort likely because of the cognitive impact of the stroke lesion.

Manual GCA ratings are time-consuming taking around three minutes on average for our highly experienced rater and results vary even across experts. In contrast, our automated DL-tool measured GCA score in only four seconds, easily and reliably and could be applied at scale in large pragmatic studies. In addition, translation into the clinic has the potential for substantial patient benefit. A clinically approved tool at point of care would generate a GCA score for any patient undergoing a CT head scan for any reason facilitating radiologists' workflows and supplementing standard radiology reports. The numeric form of the GCA output would also enable imaging variables to be combined with clinical measures in (automated) EHR risk prediction algorithms, for example to predict future dementia, delirium, falls and functional decline (22, 23).

The strengths of our study include the use of large real-world CT-brain scans from different scanners in representative older patient cohorts to validate our state-of-the-art image analysis tool providing outputs in the form of a clinically relevant GCA score, easily understood by clinicians and suitable for inclusion in algorithms. The tool will enable large-scale extraction of cerebral atrophy, a key brain frailty marker, exploiting information that is currently under-used. There are also some limitations. First, although our training and testing dataset was reasonably large, the tool could be further optimised by adding other large, labelled datasets with varying case-mix, ethnicities and geographical origin. Second, the tool tended to underestimate high GCA scores and overestimate lower GCA scores, which may be a result of the training set containing relatively few scans at the extremes. Third, the model was trained against visual ratings from a single rater and the model will therefore likely reflect any individual systematic errors. Future work will include tool training on scans from different geographic regions enriched for scans without/with very severe atrophy, comparison with MRI-measures and measurement of relative temporal lobe atrophy, a key biomarker of Alzheimer's disease.

In summary, our DL-based CT tool reliably measures global brain atrophy, in the form of a numeric GCA score, automatically and without requiring user input. The DL-tool performed well versus expert human ratings across a wide range of GCA scores obtained from real world CT-brain scans from older patients, suggesting clinical utility. Our tool will facilitate the extraction of brain atrophy data at scale for use in research, and with eventual clinical adoption will provide real-time atrophy measurements to support dementia and brain frailty diagnosis, monitoring and risk prediction.

**Table 1**: Clinical and Demographic data of participants

|  | ORCHARD-EPR (n= 578) | OCS (n=200) |
|---|---|---|
| Age mean (SD) | 81/8 | 78/8.4 |
| Sex (male/female) | 285/293 | 88/112 |
| Cognitive impairment (AMT<9) | 335 | - |
| Median/IQR AMTS score | 6/4 | - |
| Mean/SD number of OCS tasks impaired | - | 11.9/4.6 |
| Mean/SD GCA score | 18.7/5.8 | 11.1/6.4 |

**Figure legends**

**Figure 1**: Histograms for the combined dataset for (A) age (ORCHARD-EPR and OCS cohorts) (B) GCA score distribution obtained using visual ratings by EC (all three patient cohorts).

**Figure 2.** Scatter plots showing DL tool predicted GCA scores vs GCA scores from rater-1 showing a linear trend for (A) Overall dataset (B) ORCHARD-EPR (C) OCS (D) legacy scans (E) GCA scores from rater 1 vs GCA scores from rater 2 (F) GCA scores from rater 2 vs DL tool predicted GCA scores.

**Figure 3**. Bland-Altman plot indicating the errors in GCA score predictions for the DL tool vs rater-1 for (A) all patients, (B) ORCHARD-EPR (C) OCS and (D) legacy scans.

**Figure 4**. Normalised confusion matrix demonstrating the accuracy of classification into three classes i.e. mild atrophy (GCA : 0-11), moderate atrophy (GCA : 12-21) , and severe atrophy (GCA 22-39) for (A) DL tool vs rater 1 (B) DL tool vs rater 2 (C) rater 1 vs rater 2.

**Figure 5**. For ORCHARD-EPR (A) Age vs GCA score from DL-tool (B) AMTS vs GCA score from DL-tool (C) box plot of GCA score from DL tool for age <=75 vs >75 (D) box plot of GCA score from DL-tool for cognitively impaired and normal patients.

**Figure 6**. For OCS (A) Age vs GCA score from DL-tool (B) proportion subtasks impaired score vs GCA score from DL-tool (C) box plot of GCA score from DL tool for age <=75 vs >75 (D) box plot of GCA score from DL-tool for cognitively unimpaired and imapaired patients.

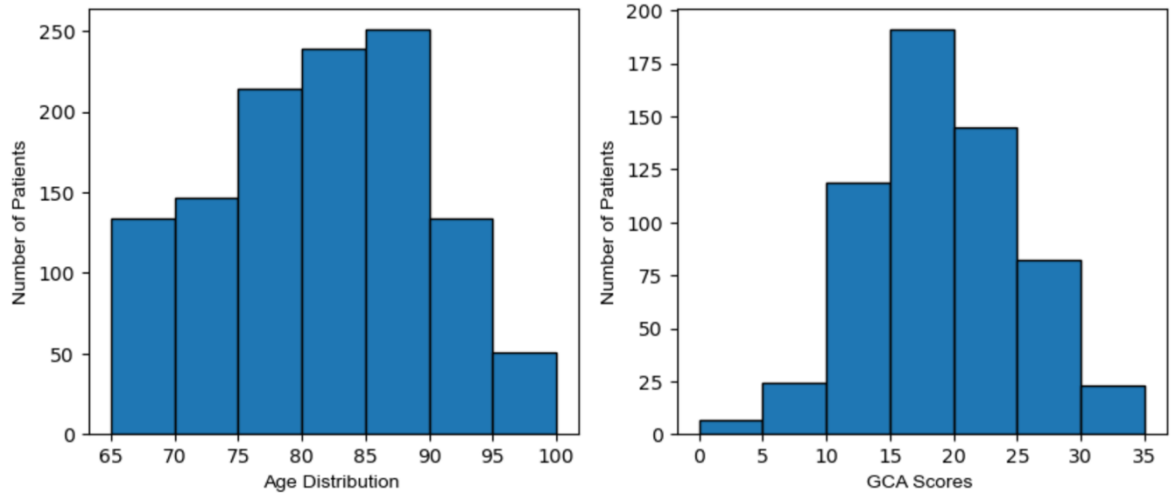

**Figure 1**: Histograms for the combined dataset for (A) age (ORCHARD-EPR and OCS cohorts) (B) GCA score distribution obtained using visual ratings by EC (all three patient cohorts).

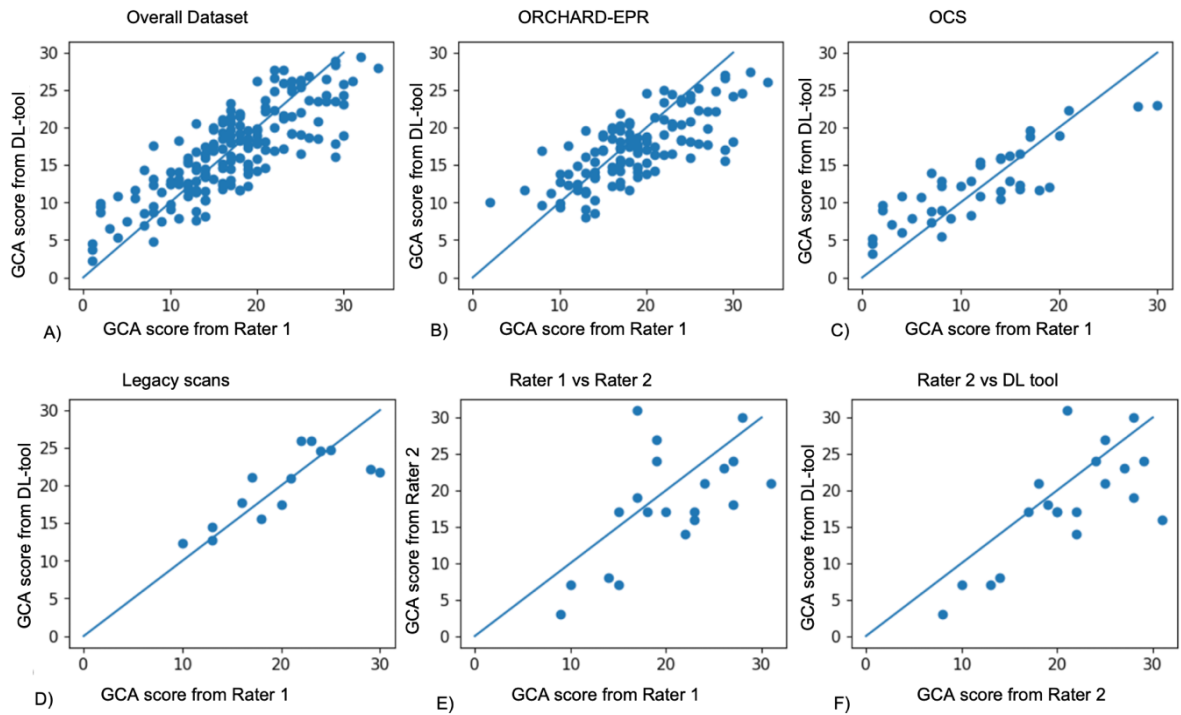

**Figure 2.** Scatter plots showing DL tool predicted GCA scores vs GCA scores from rater-1 showing a linear trend for (A) Overall dataset (B) ORCHARD-EPR (C) OCS (D) legacy scans (E) GCA scores from rater 1 vs GCA scores from rater 2 (F) GCA scores from rater 2 vs DL tool predicted GCA scores.

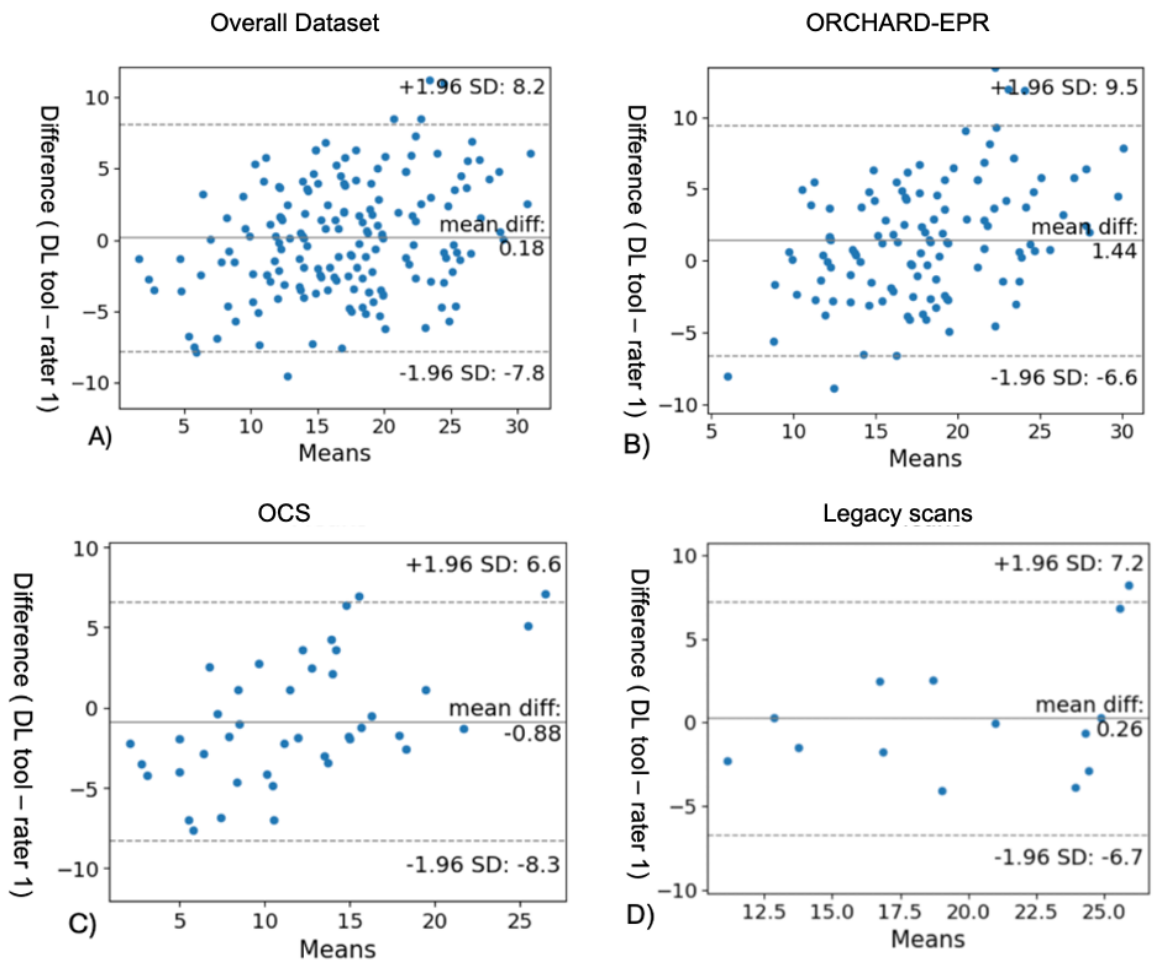

**Figure 3**. Bland-Altman plot indicating the errors in GCA score predictions for the DL tool vs rater-1 for (A) all patients, (B) ORCHARD-EPR (C) OCS and (D) legacy scans.

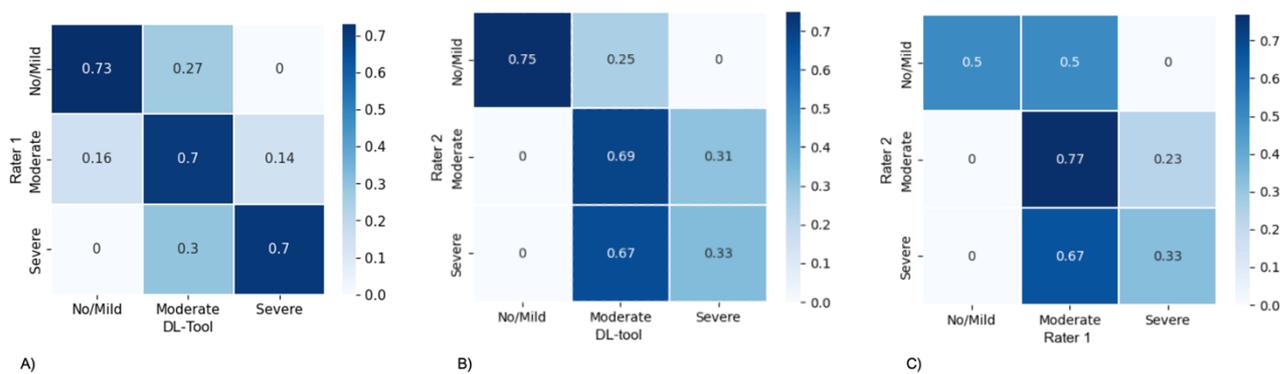

**Figure 4**. Normalised confusion matrix demonstrating the accuracy of classification into three classes i.e. mild atrophy (GCA : 0-11), moderate atrophy (GCA : 12-21) , and severe atrophy (GCA 22-39) for (A) DL tool vs rater 1 (B) DL tool vs rater 2 (C) rater 1 vs rater 2.

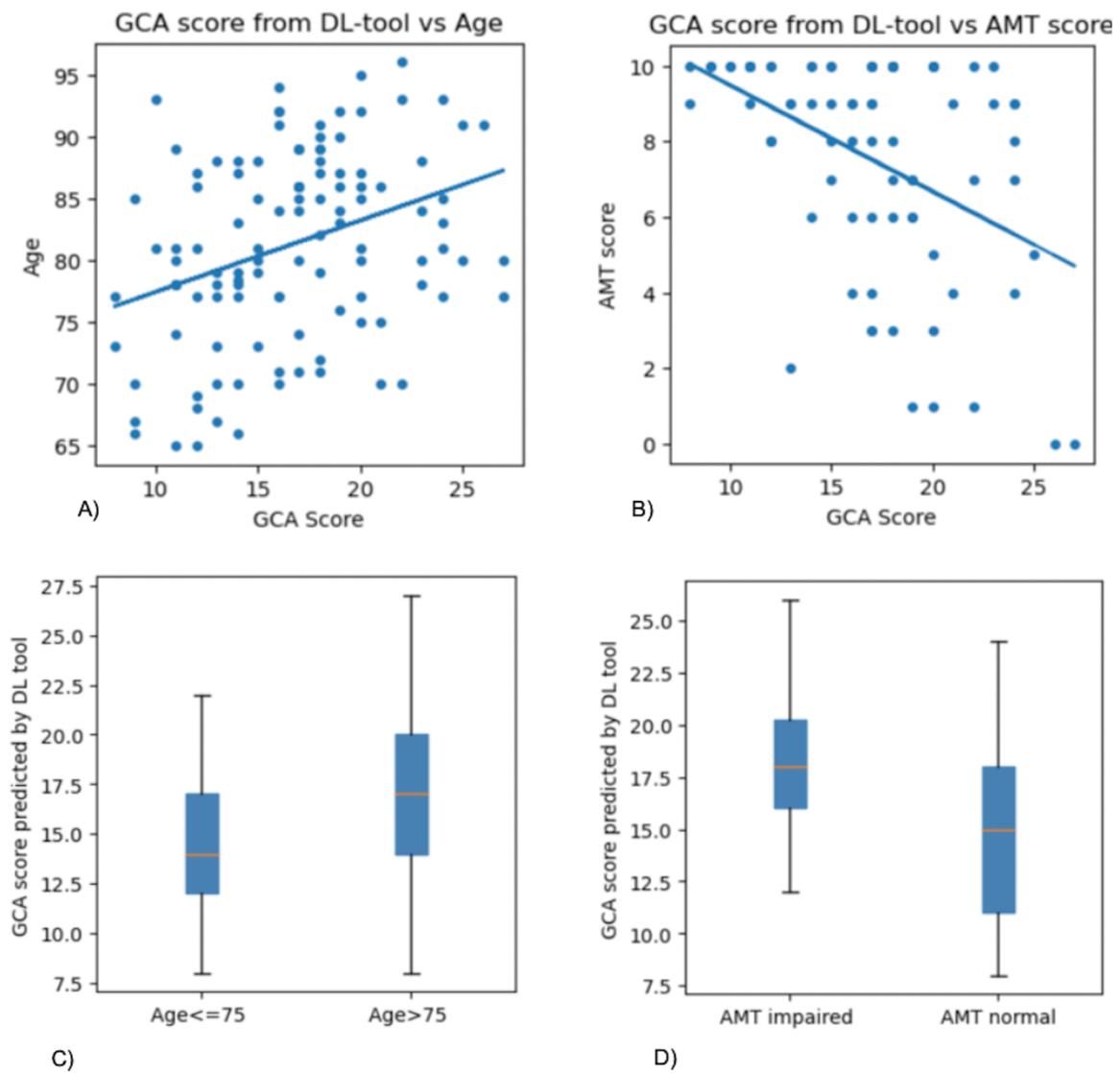

**Figure 5**. For ORCHARD-EPR (A) Age vs GCA score from DL-tool (B) AMTS vs GCA score from DL-tool (C) box plot of GCA score from DL tool for age <=75 vs >75 (D) box plot of GCA score from DL-tool for cognitively impaired and normal patients.

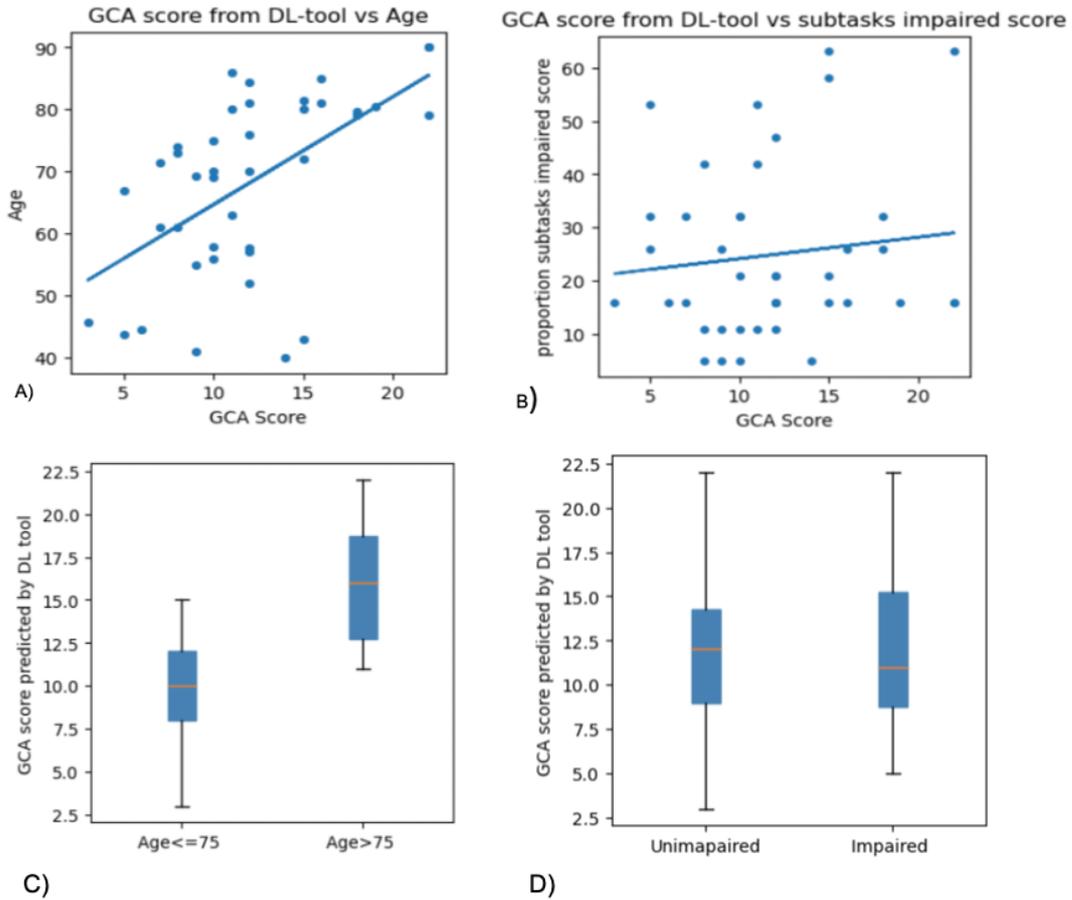

**Figure 6**. For OCS (A) Age vs GCA score from DL-tool (B) proportion subtasks impaired score vs GCA score from DL-tool (C) box plot of GCA score from DL tool for age <=75 vs >75 (D) box plot of GCA score from DL-tool for cognitively unimpaired and impaired patients.

**Supplementary Methods**

**Operationalisation of Global Cortical Atrophy (GCA) scoring as applied to CT-brain scans**

Background

Although the original Global Cortical Atrophy (GCA) paper describes the assessment of atrophy across 13 brain regions in both hemispheres,[1] the application of the scoring system in practice is open to interpretation. We therefore developed a system to operationalise application of the GCA score to axial slices on CT-brain scans acquired as part of standard clinical care and assessed inter-and intra-rater reliability in GCA scores between two raters.

Atrophy is assessed based on the amount of cerebrospinal fluid (CSF) visible in brain imaging compared to a baseline measure for the region. The baseline can be approximated through assessment of prior brain scans of the individual, or brain scans of young adults without cortical atrophy. CSF collects in areas where there is void space and appears darker than surrounding tissue on CT-brain scans. The detailed scoring methods to operationalise the GCA score were developed through discussion between the two raters and meetings with a third researcher with previous extensive experience in validating the GCA scale on both CT- and MRI-brain scans[2]. Both raters used the same baseline reference CT-brain scans of a 25-year-old male and 30-year-old female for GCA=0.

Sulcal dilation evaluation in general

GCA=0: Sulci are slightly visible compared to the surrounding brain tissue, and thin with tapered ends. They do not extend into white matter. When the slices are viewed moving along the vertical body plane, the visible sulci do not extend over a large area proportional to the brain volume and number of slices in the scan. The brain is flush against the skull, with no space between the brain and skull. Some space anteriorly and posteriorly at the longitudinal cerebral fissure is normal. Normal sulci are more visible and defined in the superior region of the brain of axial slices.

GCA=1: Widened sulci appear to have rounded openings at the skull and blunted ends, extend deeper into the brain, and appear more frequently over the region.

GCA=2: Volume loss of the gyri is seen when the brain is no longer flush with the skull, the space between the skull and brain is more contrasted to surrounding brain tissue, which occurs consistently throughout most of the region.

GCA=3: Severe volume loss of a region is seen when there is a wide opaque space between the skull and brain with substantial recession, the sulci extend deep into the brain towards the longitudinal cerebral fissure, the gyri appear thin, and the sulcal space is wider than the gyri, which can present as knife-blade-atrophy.

Evaluating sulcal dilation according to region

The original GCA scoring criteria do not elaborate on the regional presentation of none, mild, moderate, or severe sulcal widening. Owing to the variability in individual brain scan presentation, the following scoring criteria has been developed to define the *most likely* score based on the number of features present for each region and score as there will inherently be scans with regional features of more than one score. For example, if a brain scan region demonstrates defined features of a 1 and 2, whichever score features are most prevalent would be the recommended score.

*Frontal Lobe*

GCA=0: No space between the frontal lobe and skull, except for a small amount of CSF in the longitudinal cerebral fissure. Sulci are faint with tapered ends, and do not extend into the white matter. The visible sulci do not span over a large portion of the brain when scrolling through the axial imaging slices along the vertical body plane. Superiorly, sulci are visible but narrow.

GCA=1: Sulci are slightly widened indicated by darkening and blunting of the ends and elongation. Rounding of sulcal entrance may be visible and the sulcal edges are more defined. Widening can be seen in the longitudinal cerebral fissure where the space is consistently contrasted to surrounding brain tissue. Sulcal widening can also be seen anteriorly and/or laterally throughout the region.

GCA=2: Volume loss of the gyri is seen with CSF consistently visible between the skull and brain. The sulci are very dark and widened to the extent that some volume loss of the gyri is visible. The dark sulci extend centrally into the brain beyond what is anatomically normal. The longitudinal cerebral fissure is widened throughout its length with dark space visible as the imaging is observed scrolling through axial slices. .

GCA=3: Severe volume loss of the gyri is seen with large areas of visible CSF. Significant recession of the frontal lobe from the anterior skull is seen. The dark and widened sulci extend deeply into the brain, and the CSF volume between the sulci and between the brain and skull is greater than the gyri volume of the frontal lobe. The gyri appear significantly atrophied which may be seen as knife blade atrophy.

*Parieto-occipital Lobe*

GCA=0: The longitudinal cerebral fissure is normally wider than other sulci. Otherwise, no CSF is visible between the skull and parietal or occipital lobes. Some faint sulci are visible posteriorly and superiorly.

GCA=1: Sulcal widening is indicated by increased CSF within the sulci, and more frequently visible sulci laterally and posteriorly. Widened sulci are consistently visible as the scan is observed by scrolling through axial slices. Sulcal widening can be seen at the longitudinal cerebral fissure.

GCA=2: Volume loss is indicated when CSF is consistently visible between the skull and parieto-occipital lobe. The gyri have atrophied, and the sulci extend deeply in the white matter. Volume loss can be seen centrally about the longitudinal cerebral fissure with CSF consistently visible.

GCA=3: Severe volume loss is visible where the CSF between the skull and parieto-occipital lobe is opaque, wide, and consistently visible vertically and laterally throughout the region. The gyri have lost significant volume and are recessed from the skull. Volume loss can be seen about the longitudinal cerebral fissure. The volume of CSF in the sulci may be greater than the gyri volume.

*Temporal Lobe*

GCA=0: The sylvian fissure is thin with a tapered end, and does not extend deeply into the white matter. It is normal for the opening to appear slightly widened. There are few sulci visible laterally in the region.

GCA=1: The sylvian fissure is dark and slightly dilated. It extends into the brain, with a more contrasted and/or rounded opening, with no loss of gyri volume. Some contrasted sulci are seen laterally.

GCA=2: The sylvian fissure is opaque and thick. It may be elongated into the brain and the gyri surrounding the sylvian fissure opening are recessed. The widened sylvian fissure is visible over a larger region as the scan is moved along the vertical body axis. The lateral temporal lobe is slightly recessed from the skull with visible CSF throughout the region.

GCA=3: The lateral temporal lobe is constantly and significantly recessed from the skull. There is significant volume loss of the gyri around the sylvian fissure which is widened and extends deeply into the brain.

Ventricular dilatation

There is less challenge in visually rating ventricular dilation on axial slices. The ventricular shape and size is usually easily seen and quantified in comparison to regional quantification of sulcal dilation.

Frontal Horn

GCA=0: The frontal horn is thin with tapered ends.

GCA=1: The frontal horn is widened and darkened. It extends outwards, increasing the frontal horn width. The ends are slightly rounded and defined.

GCA=2: The frontal horn extends outwards, and the angle becomes less defined.

GCA=3: The frontal horn is thick; the angle is minimally defined, and the ends are bulbous. The frontal horn width is greatly increased.

Occipital Horn

GCA=0: The occipital horn is small, speck-like, and slightly contrasted to the surrounding tissue. The edges are undefined/blurred and tapered. It is not visible over many slices proportional to the brain volume.

GCA=1: The occipital horn is widened, with increased opacity, and rounded edges resembling a button. The occipital horn is visible over more slices relative to the brain volume.

GCA=2: The occipital horn is large and opaque with volume loss of surrounding tissue. The edges are clearly defined with a pronounced contrast between the CSF and brain tissue.

GCA=3: The occipital horn takes up a large portion of the occipital region and may follow the contour of the skull.

Temporal Horn

The temporal horn varies greatly depending on scan slice thickness and features captured in the slice.

GCA=0: The temporal horn is a faint, low-contrast pinpoint which may have a thin tail depending on scan slice location. It is only visible on one or two slices and appears as if it were drawn with a pencil.

GCA=1: The temporal horn is slightly darker and widened as though it was drawn with a crayon. It is more likely that a tail is visible.

GCA=2: The temporal horn is large or elongated, with higher contrast and potentially opaque centre, as if it was drawn with a marker Evidence of the enlarged temporal horn may span over more than one or two slices.

GCA=3: The temporal horn is very opaque and large; it may span over three or more slices as if it were drawn with a thick-tip marker. There is volume loss of surrounding tissue owing to ventricular dilation.

Third Ventricle

GCA=0: The third ventricle is skinny with the width of a pencil body and is only visible over a few slices.

GCA=1: The third ventricle is slightly widened, and the edges are less defined, like a crayon.

GCA=2: The third ventricle is widened with a thick appearance like a marker.

GCA=3: The third ventricle is severely widened with rounded edges and bulbous ends, like a highlighter and spans over many slices.

Other factors and clinical relevance

Factors such as scan angle, artifacts, movement, hyperintensities, scan noise, or pathologies may affect the rating process. Scan angle may require the rater to conduct mental spatial manipulation to account for how the angle may affect the appearance of the sulci or ventricles. For example, a normal sulcus sliced on an angle along its widest plane may appear wider than it is. Pathology such as acute stroke and oedema, and scan noise may obscure a region making rating difficult or impossible. If there is an artifact or pathology preventing the rating of a region in one hemisphere, the affected region is given the same score as the anatomically homologous region in the opposite hemisphere. When both hemispheres are affected, the region may not be assessable.

The method outlined in this protocol was previously tested on CT-brain scans from a stroke cohort.  Each patient had a single CT and MRI conducted on average 2.41 days apart.[2] The results of this study showed that trained raters using the above method to operationalise the GCA scoring system achieve high rates of inter- and intra-rater reliability.

**Supplementary Results**

**Time taken to complete GCA score**

The time taken to rate 18 scans with the GCA score was recorded by rater-1. The results are displayed in the Table below. The average time take was just under 3 minutes (2 minute and 56 seconds). This was by a highly experienced rater who had rated hundreds of scans and would likely take longer for someone less familiar with apply the scoring system.

**Supplementary Table 1**. Time taken to visually rate the GCA score by rater-1 for a sample of 18 consecutive scans.

| Scan number | Modality available | Modality rated | Time taken to visually rate the GCA score in seconds |
|---|---|---|---|
| 1 | CT + MR | CT | 101 |
| 2 | CT only | CT | 175 |
| 3 | CT + MR | CT | 185 |
| 4 | CT only | CT | 179 |
| 5 | CT only | CT | 145 |
| 6 | CT only | CT | 217 |
| 7 | CT only | CT | 176 |
| 8 | CT only | CT | 165 |
| 9 | CT + MR | CT | 244 |
| 10 | CT + MR | CT | 180 |
| 11 | CT only | CT | 142 |
| 12 | CT only | CT | 247 |
| 13 | CT only | CT | 162 |
| 14 | CT only | CT | 106 |
| 15 | CT only | CT | 220 |
| 16 | CT only | CT | 223 |
| 17 | CT only | CT | 119 |
| 18 | CT only | CT | 173 |

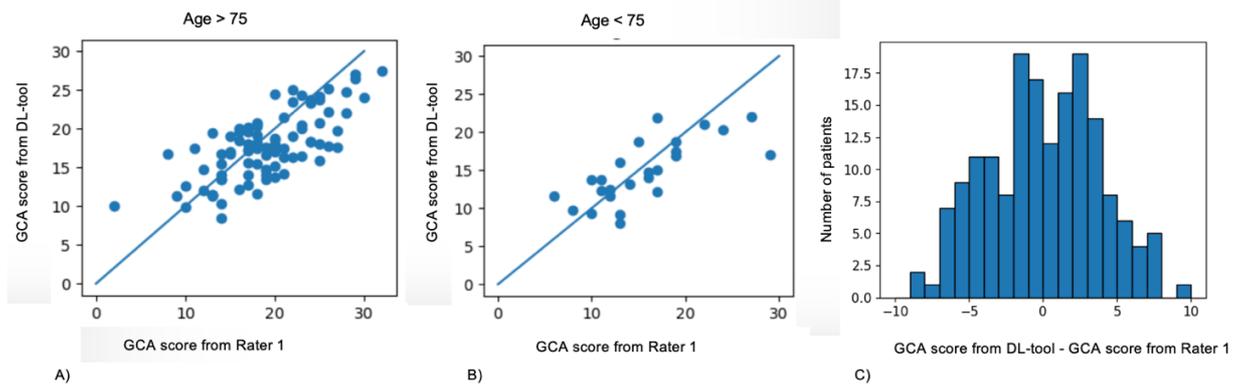

**Supplementary Figure 1:** Scatter plots showing DL tool predicted GCA scores vs GCA scores from visual ratings (rater-1) showing a linear trend for (A) Age >75 (B) Age <75 (C) Error histogram demonstrating the frequency of errors in GCA predictions.